# Asymmetric planar terahertz metamaterials

Ranjan Singh,[1,2,*] Ibraheem A. I. Al-Naib,[3] Martin Koch,[3] and Weili Zhang[1]

[1]*School of Electrical and Computer Engineering, Oklahoma State University, Stillwater, Oklahoma 74078, USA*
[2]*Center for Integrated Nanotechnologies, Materials Physics and Applications Division, Los Alamos National Laboratory, Los Alamos, New Mexico 87545, USA*
[3]*Physics Department, Philipps-Universität Marburg, Renthof 5, 35032 Marburg, Germany*
*\* ranjan@lanl.gov*

**Abstract:** We report an experimental observation of three distinct resonances in split ring resonators (SRRs) for both vertical and horizontal electric field polarizations at normal incidence by use of terahertz time-domain spectroscopy. Breaking the symmetry in SRRs by gradually displacing the capacitive gap from the centre towards the corner of the ring allows for an 85% modulation of the fundamental inductive-capacitive resonance. Increasing asymmetry leads to the evolution of an otherwise inaccessible high quality factor electric quadrupole resonance that can be exploited for bio-sensing applications in the terahertz region.

©2010 Optical Society of America

## 1. Introduction

In electromagnetism, a metamaterial (MM) is an artificially designed material that gains effective properties from its structures rather than inheriting them directly from the materials it is composed of. There has been rapid progress in study of MMs in the last decade after its constituent subwavelength building block split ring resonators (SRRs) were first proposed by Pendry et al. and then extensively studied by several groups [1-11]. The SRRs allow for a control and manipulation of light on subwavelength length scales which is very important goal for incorporating optics into micro and nanotechnology. These artificially engineered resonators have found a broad range of applications such as absorbers, antenna structures, frequency selective surfaces, sensors, high frequency modulators, and composite materials [9-20]. Since its inception, the SRRs have undergone several miniaturized refinements in an attempt to optimize the resonances supported in the structure as these resonances hold the key in tailoring the effective permeability and permittivity of the MMs. At the fundamental inductive-capacitive (LC) resonance mode, the SRRs provide a strong magnetic response with negative permeability required for achieving a negative index of refraction. In order to achieve this, both the SRR and the wave propagation direction should lie in the same plane so that the induced circular current creates a magnetic moment parallel to the magnetic field of the incident wave. For normal wave incidence the magnetic moments associated with the induced currents are perpendicular to the plane of the SRR array and, therefore, the interaction of the induced magnetic moments with the magnetic field of the incident wave is negligible. In most of the previous works at terahertz frequencies, the focus has been on exciting the LC resonance and the dipole resonance for horizontally polarized electric field along the gap arm of SRR at normal incidence [21-28].

Here we demonstrate the excitation of three distinct resonance modes in SRRs for both horizontal and vertical incident electric field polarizations by introducing asymmetry in their structure. The three resonance modes are the LC, dipole and the quadrupole resonances. Quadrupole mode is known as a weak mode [21]. However, we show that a strong one can be achieved by crossing the symmetry of the structure. When the incident field is horizontal and parallel to the SRR gap, the LC resonance becomes weak but a strong evolution of quadrupole resonance is witnessed with increasing asymmetry. For the orthogonal incident polarization, we observe the simultaneous formation of a weak LC resonance and a strong quadrupole resonance as the degree of asymmetry is increased in a step wise fashion by displacing the split ring gap from the center towards the extreme corner of the gap arm in the SRR. An extremely sharp quadrupole resonance with a high quality factor ($Q$) is observed experimentally for both incident polarizations. Such narrow resonances can be exploited for highly efficient sensing and frequency selection in the terahertz domain.

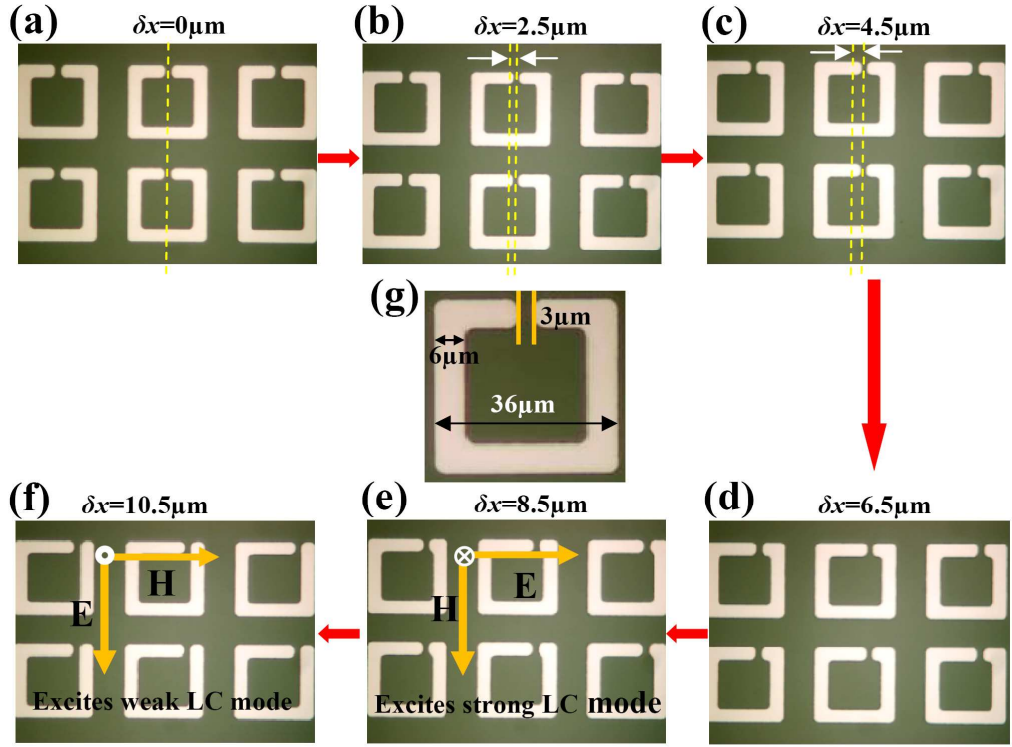

Fig. 1. (a) - (f) is the microscopic image of the six sample arrays; the gap in the SRR is moved in a step wise fashion by different values of '$\delta x$' from the center (g) Unit cell with dimension parameters, $t = 6$ μm, $d = 3$ μm, $l = 36$ μm, Al metal film thickness is 200 nm. The periodicity of the unit cells in all samples is $P = 50$ μm

## 2. Experiment

We employ broadband terahertz time-domain spectroscopy (THz-TDS) was employed to characterize the asymmetrical planar metamaterials [29]. The photoconductive-switch based THz-TDS system consists of four parabolic mirrors configured in an 8-F confocal geometry that enables a 3.5 mm diameter frequency independent beam waist for small sample characterization. Six sets of planar SRR metamaterials with 200 nm thick Al metal structures are fabricated by conventional photolithography on a silicon substrate (0.64-mm-thick, n-type resistivity 12 Ω cm). Asymmetry in the SRRs is introduced by displacing the gap gradually from the center, as shown in Figs. 1(a)-1(f) with $\delta x = 0, 2.5, 4.5, 6.5, 8.5$, and $10.5$ μm, where '$\delta x$' represents the gap displacement parameter. Figure 1(g) shows the diagram of a SRR unit cell with a minimum feature $d = 3$ μm in the splits of the rings and other dimensions $t = 6$ μm,

$l$ = 36 µm, and the lattice constant $P$ = 50 µm. Each SRR array has a 1 cm × 1 cm clear aperture. In the first case the orientation of the incident terahertz field is perpendicular to the SRR gap which excites the weak LC resonance. In the second case the field is along the SRR gap in order to excite the regular LC and the dipole mode resonances. In the THz- TDS measurements, each metamaterial sample is placed midway between the transmitter and receiver modules in the far-field at the focused beam waist and the terahertz waves penetrate the SRRs at normal incidence.

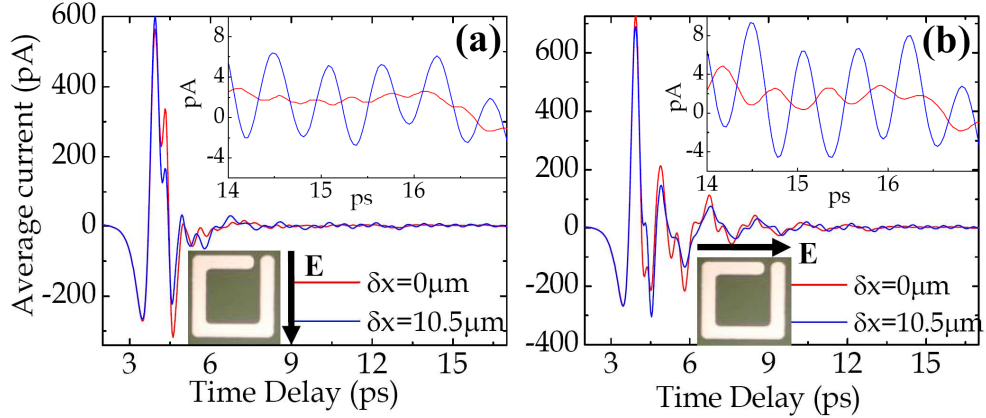

Fig. 2. (a) – (b) Measured sub-picosecond transmitted pulses through the symmetric ($\delta x$ = 0 µm) and extreme asymmetric ($\delta x$ = 10.5 µm) MM samples with electric field vertical and horizontal to the SRR gap respectively. The inset shows the blow up of the later time delay pulses.

The time domain data were taken for all the six MM samples in a sequential order for both orientations of the electric terahertz field. Figure 2(a) shows the measured terahertz pulses transmitted through the symmetric SRR ($\delta x$ = 0 µm) and the extreme asymmetric SRR ($\delta x$ = 10.5 µm) while the terahertz field is vertical to the SRR gap. For asymmetric SRR, a strong ringing in the pulse is observed for late times as shown in the top inset of Fig. 2(a). Figure 2(b) shows the pulse measured for the other orientation when the field is along the SRR gap. We see an increase in the pulse oscillation towards later times for the asymmetric sample. As we will see below that this is due to strengthening of a quadrupole resonance at around 1.72 THz.

## 3. Measurement and Simulation

The transmission is extracted from the ratio of the Fourier-transformed amplitude spectra of the samples to the reference, defined as $|E_s(\omega)/E_r(\omega)|$, where $E_s(\omega)$ and $E_r(\omega)$ are Fourier-transformed time traces of the transmitted electric fields of the signal and the reference pulses, respectively. Figure 3(a) shows the measured transmission spectra through all the six samples with varying asymmetry. The field orientation with respect to the SRR gap is shown in the inset. For the perfect symmetric SRR when the gap is right at the center of the arm ($\delta x$ = 0 µm), there is excitation of only a typical dipole resonance at 1.36 THz due to linear oscillating currents in the SRR arms parallel to the incident field. Gradually, as we introduce asymmetry in the structure by altering the gap displacement, we observe the evolution of a weak LC resonance at 0.55 THz where the transmission gradually changes from 100% to 75%. The formation of the resonance is most pronounced for $\delta x$ = 10.5 µm when the SRR gap is pushed all the way to the extreme corner. The $Q$-factor for the LC resonance of this sample is 7. The increase in asymmetry of the MM also gives rise to another sharp quadrupole resonance feature with *Q*-factor as high as 35 at 1.72 THz for the SRR with $\delta x$ = 10.5. The dipole resonance feature blue shifts by 160 GHz for extreme asymmetry. Figure 3(b) shows the simulation results for identical experimental conditions and most of the transmission spectra are found in good agreement with the measurements [30].

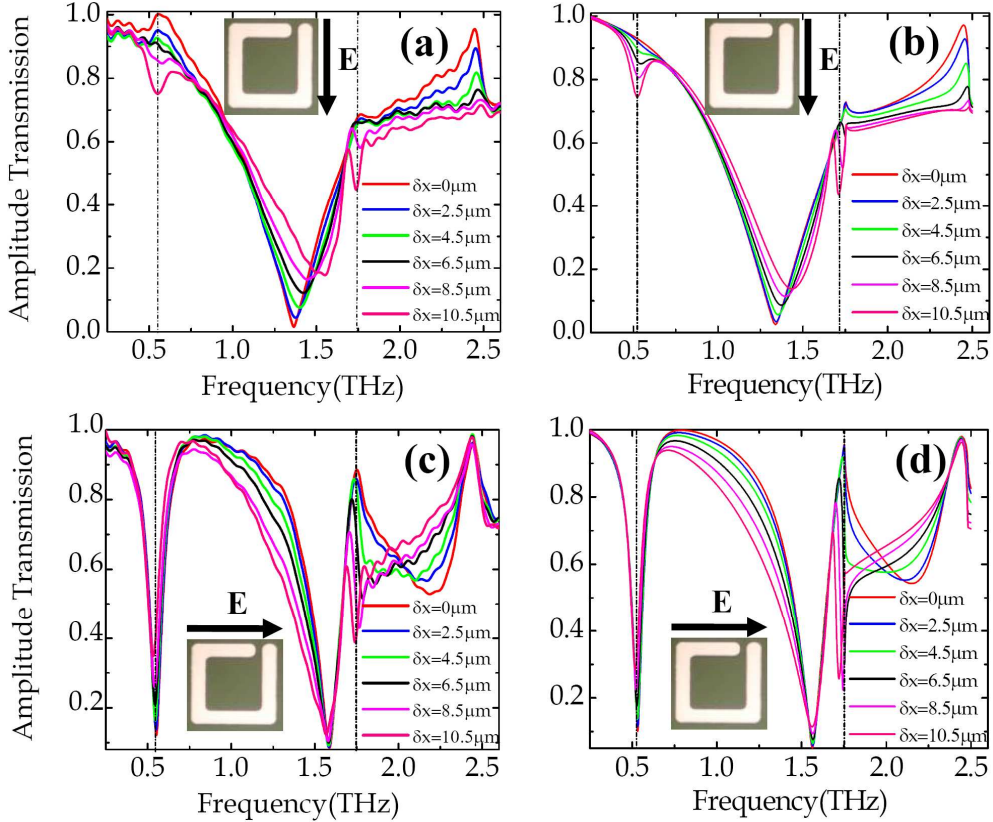

Fig. 3. Measured (a) and simulated (b) amplitude transmission spectra of with varying asymmetry in SRR for vertical E field polarization. (c), (d) show the measured and simulated spectra for horizontal polarization.

As the incident terahertz field is aligned along the SRR gap, it usually excites the two regular resonance modes, the lower frequency LC resonance and the higher frequency dipole resonance. Figure 3(c) reveals the change in both these regular resonance modes as the SRR gap is swept along the arm. The LC resonance undergoes slight shrinking in the line width and the transmission at the resonance frequency gets enhanced from 12% ($\delta x$ = 0 µm) to almost up to 35% ($\delta x$ = 10.5 µm). The dipole resonance does not shift in frequency but undergoes significant broadening by about 15 GHz. As the asymmetry is increased, there is appearance of another sharp quadrupole resonance feature at 1.72 THz. Thus the weak LC mode resonance for vertical polarization and the high *Q* quadrupole resonance features at 1.72 THz for both orientations appear entirely due to the introduction of asymmetry in the SRRs. It should be noted that the weak LC mode resonance and the regular strong LC resonance occurs at the same frequency but for orthogonal electric field polarizations. Figure 3(d) represents the simulation for the horizontal polarization and it reproduces all of the transmission spectrum features identical with the measurement.

## 4. Discussion

In order to get an insight into the nature of resonances for the asymmetric SRRs, we simulate the structures using the Microwave Studio CST frequency domain solver [30] and focus at the surface current profile. Figures 4(a)-(c) depict the current distribution at the aforementioned three resonances, i.e. the weak LC, the dipole, and the quadrupole resonance. The arrows indicate instantaneous directions of the current flow at certain phase shifts. At 0.52 THz the current flows in the circular direction throughout the resonator. For vertical incidence polarization as soon as the gap is displaced from the center of the SRR, the symmetry is broken and the asymmetry of the SRR with respect to the external electric field **E** leads to a charge distribution asymmetry, which is compensated through a circular current. Thus, the electric field of the incoming wave can couple to the resonance of the circulating currents in the SRR if there is symmetry breaking involved and the intensity of the current increases as the asymmetry in the SRR is increased by displacing the capacitive gap from the central position. The strength of the LC resonance is minimum for the minimum asymmetry at $\delta x = 0$ µm and is maximized for the most asymmetric case at $\delta y = 0$ µm, as shown in Fig. 5(a) where the incident electric field is fixed but the gap is translated first in the $y$ direction and then in the $x$ direction. The transmission at the LC resonance reveals that if we could actively move the gap along the SRR arm from the center of the right arm ($\delta y = 0$) to the center of the top

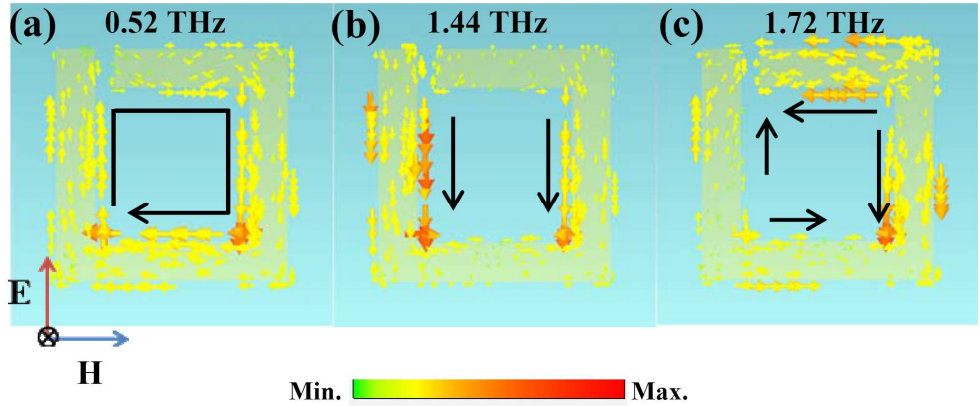

Fig. 4. Simulated surface current distribution of (a) weak LC mode resonance at 0.52 THz , (b) dipole resonance at 1.44 THz, and (c) quadrupole resonance at 1.72 THz for vertical E field polarization for asymmetry parameter $\delta x$= 10.5 µm

arm ($\delta x = 0$), then a modulation of 85% could be easily achieved. The role of $\delta y$ displacement of the gap is only to decrease the degree of asymmetry in SRR with respect to the incident electric field. This allows for a much higher modulation at the LC resonance. The second resonance at 1.44 THz is attributed to the excitation of dipole like plasmons in the vertical arms of the SRRs. These currents are parallel to the polarization of the electric field. This electric resonance is related to the plasmon resonance in a thin continuous wire but shifted to nonzero frequency as a result of the additional depolarization field arising from the finite side length of the SRR. It appears independent of the excitation of the circular currents of the LC resonance and also occurs for the closed rings. This electric resonance is also referred to as the particle-plasmon or dipole resonance. The linear current distribution as seen in Fig. 4(b) radiates strongly to the free space giving rise to a broader resonance in the transmission spectrum. The dipole resonance blue shifts when the SRR gap is displaced from its symmetric position since the alteration in the nearby horizontal SRR arm modifies the radiative properties of the dipole oscillating along the incident E field direction [31]. The quadrupole resonance is observed at 1.72 THz. The four arrows show how the current distribution behaves in general. There are two sets of out of phase current distribution. Introducing the asymmetry leads to this kind of current distribution that scatters the electromagnetic field very weakly and dramatically reduces the coupling to the free space thus causing a huge reduction in the radiation losses which eventually leads to very sharp resonance. The quadrupole resonance is inaccessible in the symmetrical structure.

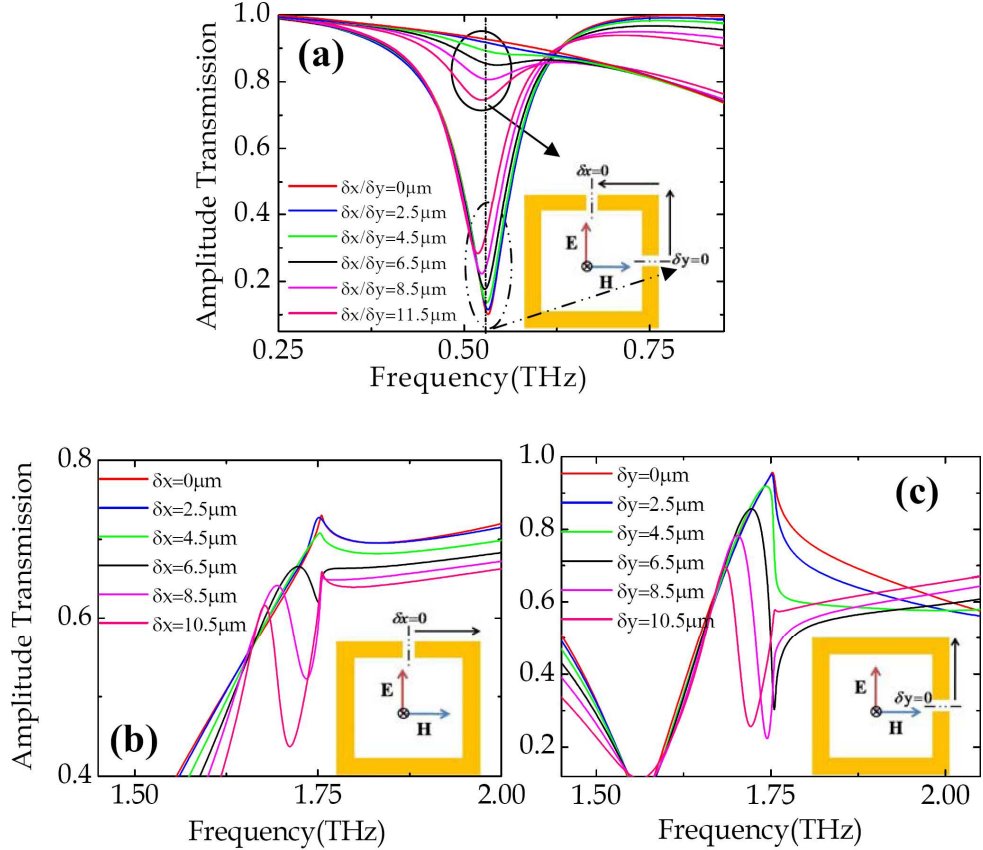

Fig. 5. Amplitude transmission at (a) LC resonance as the gap is swept from right arm of SRR to the top arm keeping the incident E field polarization fixed. Simulated zoomed quadrupole resonance transmission for (b) Vertical polarization and (c) Horizontal polarization with increasing asymmetry.

The gradual evolution of this resonance can be observed in Figs. 5(b) and 5(c) where the $Q$ factor reaches as high as 55 for vertical polarization and 93 for horizontal polarization when the SRR gap is displaced by $\delta x = \delta y = 8.5$ μm. The $Q$ factor is extracted from the simulated transmission curves. The high $Q$ factor of the electric quadrupole resonances can be easily exploited for sensing and narrow band filtering purposes.

## 5. Conclusion

In conclusion, we have characterized asymmetrical planar terahertz SRRs and excited the LC, dipole and the quadrupole mode resonances for vertical and horizontal electric field polarizations at normal incidence. Shifting the gap position allows us to engineer the transmission properties of metamaterials across a large portion of electromagnetic spectrum. A very high passive modulation of the LC resonance has been achieved by varying the asymmetry parameter. This property can find applications in the design of terahertz modulators in which the SRR gap could be actively displaced by external pumping. The symmetry breaking also gives access to the high $Q$ quadrulpole resonances where the concentrated field in a much smaller volume would open up avenues for efficient bio-sensing applications in the terahertz domain.

**Acknowledgements**

The authors thank H. T. Chen, J. F. O'Hara and J. Zhou for their support and discussions. This work was supported by the U.S. National Science Foundation.